\documentclass[aps,prd,letterpaper,showpacs,twocolumn,preprintnumbers,floatfix,superscriptaddress]{revtex4}
\usepackage{graphicx,dcolumn,bm,epsfig,cancel,hyperref}
\usepackage{amsmath}
\usepackage{amssymb, times}
\newcommand{\comment}[1]{{}}
\usepackage{color}

\begin{document}
\title{
Gravitational Wave Signals of Electroweak Phase Transition Triggered by Dark Matter
}
\author{Wei Chao}
\affiliation{Center for Advanced Quantum Studies, Department of Physics, Beijing Normal University, Beijing, 100875, China}
\email{chaowei@bnu.edu.cn}
\author{Huai-Ke Guo}
\affiliation{
CAS Key Laboratory of Theoretical Physics, Institute of Theoretical Physics, \\
Chinese Academy of Sciences, Beijing 100190, China}
\email{ghk@itp.ac.cn}
\author{Jing Shu}
\affiliation{
CAS Key Laboratory of Theoretical Physics, Institute of Theoretical Physics, \\
Chinese Academy of Sciences, Beijing 100190, China}
\affiliation{
CAS Center for Excellence in Particle Physics, Beijing 100049, China
}
\email{jshu@itp.ac.cn}


\begin{abstract}
We study in this work a scenario that the universe undergoes a two step phase transition with the first step happened to the dark matter sector and the second step being the transition between the dark matter and the electroweak vacuums,  where the barrier between the two vacuums, that is necessary for a strongly first order electroweak phase transition (EWPT) as required by the electroweak baryogenesis mechanism, arises at the tree-level. 
We illustrate this idea by working with the standard model (SM) augmented by a scalar singlet dark matter and an extra scalar singlet which mixes with the SM Higgs boson. 
We study the conditions for such  pattern of phase transition to occur and especially for the strongly first order EWPT to take place, as well as its compatibility with the basic requirements of a successful dark matter, such as observed relic density and constraints of direct detections. 
We further explore the discovery possibility of this pattern EWPT by searching for the gravitational waves generated during this process in spaced based interferometer, by showing a representative benchmark point of the parameter space that the generated gravitational waves fall within the sensitivity of eLISA, DECIGO and BBO.
\end{abstract}

\pacs{11.30.Er, 11.30.Fs, 11.30.Hv, 12.60.Fr, 31.30.jp}

\maketitle

\section{ Introduction}
%
%
The Standard Model (SM) of particle physics so far provides an excellent description of a wide 
variety of  experimental observations, nevertheless, it must be extended because it fails at least in explaining two cosmological puzzles: it does not provide a dark matter (DM) candidate and it is not possible to generate the observed baryon asymmetry of the universe (BAU).
The later one requires a strongly first order electroweak phase transition (EWPT) to provide a non-equilibrium environment if the BAU is generated via the electroweak baryogenesis mechanism~\cite{Kuzmin:1985mm,Shaposhnikov:1986jp,Shaposhnikov:1987tw,Morrissey:2012db}.
Since the 125 GeV Higgs boson is too heavy to give rise to a first order EWPT~\cite{Bochkarev:1987wf,Kajantie:1995kf}, new ingredients beyond the SM Higgs interactions may be needed.
On the other hand, how DM interacts with the SM particles remains unknown.
 The Higgs portal turns to be attractive since it may kill two birds (DM and the strongly first order EWPT~\cite{Espinosa:2011ax}) with one stone.

The characteristics of a scalar DM triggered strongly first order EWPT are twofold: (i)~It is a two-step phase transition~\cite{Cheung:2013dca,Patel:2012pi,Inoue:2015pza,Patel:2013zla,Chala:2016ykx,Vaskonen:2016yiu,Huber:2015znp} and the barrier between electroweak symmetric and broken phases arises at the tree-level, which may avoid the problem of gauge-dependence~\cite{Patel:2011th}; (ii)~The parameter space of the model is strongly constrained by the exclusion limits of direct detections and can not explain the observed relic density. 
It was pointed out in Ref.~\cite{Cline:2012hg} that, for the  SM plus a real scalar DM $S$ scenario, $S$ could only constitute up to $3\%$ of the total DM relic density, while still be relevant for the direct detection because of its sizable coupling with the SM Higgs as required by the two-step EWPT. 

In this paper, we  revisit this type of EWPT by extending the SM with a scalar singlet DM $S$ and another scalar singlet $\Phi$ which can mix with the SM-like Higgs boson. 
Due to the mixing of $\Phi$ with the SM Higgs,  there are two separate contributions to DM nucleon scattering, which may cancel with each other and leads to a negligible cross section. 
It results in a mechanism to suppress the direct detection signal and evade the currently most stringent experimental constraints from the direct detection experiments~\cite{Akerib:2016vxi,Tan:2016zwf}.
Furthermore,  the quartic interaction $\Phi^2S^2$ may contribute to the mass of $S$, which can lead to a relatively small quartic coupling of $S$ with the SM-like Higgs and result in a sizable relic density.  
As a result the tension between the DM relic density and direct detection which occurs for the SM plus singlet case~\cite{Feng:2014vea,deSimone:2014pda} can be highly loosed in this scenario.

For the thermal history of the universe, the presence of the two additional scalars allows a two step EWPT. 
This happens as follows:
As the universe cools down, $S$ gets vacuum expectation value (VEV) first and the universe transits to this phase;
As the temperature gets lower, a second minimum occurs in the SM-like Higgs and $\Phi$ subspace.
The universe then tunnels to this electroweak minimum resulting in a strongly first order EWPT with gravitational waves (GW) generated.
GW signals coming from this two-step EWPT are calculated using one benchmark point of the parameter space and our results show that the corresponding GW signals are testable in the Evolved Laser Interferometer Space Antenna (eLISA),  DECi-hertz Interferometer Gravitational wave Observatory (DECIGO), Ultimate-DECIGO and Big Bang Observer (BBO).  

The remaining of this paper is organized as follows. 
 Conventions of the model are defined in Sec. \ref{sec:model}. 
We then study the DM phenomenology in Sec.~\ref{sec:dm} and the capability of a strongly first order EWPT in Sec.~\ref{sec:ewpt}. 
The GW calculations and discovery prospects are explored in Sec.~\ref{sec:gw}, after which we present a brief summary in Sec.~\ref{sec:summary}.

\section{\label{sec:model}The model}
We present the model for two-step phase transition and scalar DM in this section.
The model extends the SM with two real scalar singlets: $S$ and $\Phi$, where $S$ is the DM candidate, while $\Phi$ gets non-zero VEV at the zero temperature and thus mixes with the SM Higgs~\footnote{Our particle settings are similar to these in Ref.~\cite{Patel:2013zla} where color breaking from two-step phase transition was proposed.}.  
The Higgs potential can be written as
\begin{eqnarray}
V_0&=&-{1\over 2 }\mu_\Phi^2 \Phi^2 + {1\over 4 }\lambda_\Phi \Phi^4  - {1\over 2 }\mu_S^2 S^2 + {1\over 4}\lambda_S^{} S^4 \nonumber \\
&& -\mu^2 H^\dagger H + \lambda ( H^\dagger H )^2+\lambda_1^{} S^2 H^\dagger H + \lambda_2^{} \Phi^2 H^\dagger H \nonumber\\
&&+\lambda_3^{} S^2 \Phi^2 , 
\label{potential}
\end{eqnarray}
where $H$ is the SM Higgs doublet.
The effective potential, critical for the EWPT, can be written as: $V_{\rm eff} =V_0+ V_{\rm CW}+V_{\rm T}$, where  $V_0$ is the tree-level potential, $V_{\rm CW}$ is known as the Coleman-Weinberg term~\cite{Coleman:1973jx}, 
$V_{\rm T} $ includes finite temperature contributions from loops~\cite{Quiros:1999jp} and bosonic 
ring~\cite{Parwani:1991gq,Gross:1980br}. 
For one-step phase transition, where Higgs potential contains no tree-level cubic term, the  barrier between the electroweak symmetric and broken phases usually 
arises from  loop corrections and 
according to the Nielsen identity~\cite{Nielsen:1975fs}, the resulting effective potential is gauge dependent. 
For the two step phase transition,  the barrier arises at the tree-level. 
So we include in the effective potential the standard one-loop $T\neq 0$ corrections, but only retain terms proportional to ${\cal O} (T^2)$ for the consideration of gauge invariance. 
Thermal masses of scalars take the following forms:
\begin{eqnarray}
\Pi_{h}&=& \left\{ {3 g^2 +g^{\prime 2 } \over 16} + {\lambda \over 2 } + {h_t \over 4 } + {\lambda_1+\lambda_2 \over 12 } \right\} T^2 ,  \\
\Pi_{s}&=&\left\{{\lambda_S \over 4}  + {\lambda_1 \over 3} + {\lambda_3 \over 6}  \right\}T^2 ,  \\
\Pi_{\phi} &=& \left\{ {\lambda_\Phi \over 4} + {\lambda_2\over 3} + {\lambda_3\over 6 } \right\} T^2   \; .
\end{eqnarray}
We require $S$ to have zero VEV at $T=0$ for the stability of the DM and parametrize the other two fields by 
$H = (0, (v_{\text{EW}} + h)/\sqrt{2})$ and $\phi = v_{\Phi} + \phi$ with 
here $v_{\text{EW}} \approx 246~\text{GeV}$ and $h_t \approx 1$. 
Minimization conditions around $(v_{\text{EW}}, v_{\Phi}, 0)$
in the $(h,\phi,s)$ space allow us to trade the two VEVs $v_{\text{EW}}$ and $v_{\Phi}$ for $\mu^2$ and 
$\mu^2_{\Phi}$ by
\begin{eqnarray}
\mu^2 = \lambda_h v_{\text{EW}}^2 + \lambda_{2} v_{\Phi}^2, \quad
\mu_{\Phi}^2 = \lambda_2 v_{\text{EW}}^2 + \lambda_{\Phi} v_{\Phi}^2 \; .
\end{eqnarray}
We also replace $\mu_S^2$ by the physical DM mass $m_S^2$,
\begin{eqnarray}
  \mu_S^2 = \lambda_1 v_{\text{EW}}^2 + 2 \lambda_3 v_{\Phi}^2 - m_S^2. 
\end{eqnarray}
With these substitutions, the mass matrix for $(h,\phi)$ is then given by 
\begin{eqnarray}
  \mathcal{M}^2 = 
  2 \left(
  \begin{array}{c c}
    v_{\text{EW}}^2 \lambda          & v_{\text{EW}} v_{\Phi} \lambda_2 \\
    v_{\text{EW}} v_{\Phi} \lambda_2 & v_{\Phi}^2 \lambda_{\Phi} 
  \end{array}
\right),
\end{eqnarray}
which can be diagonalized by a $2\times 2$ orthogonal matrix parametrized by a rotation angle
$\theta$.
The mass eigenstate $( \hat h, \hat s)$ can then be written as  
\begin{eqnarray}
    \hat h =\ \ \  c_{\theta} h + s_{\theta} \phi, 
    \quad \quad
    \hat \phi=-s_{\theta} h + c_{\theta} \phi. 
\end{eqnarray}
where $c_\theta =\cos \theta $ and $s_\theta=\sin \theta$.
We define $\hat h$ as the SM-like Higgs boson and therefore when $\theta \rightarrow0$,  $\hat h$ is purely the SM Higgs.
With this definition, three of the parameters in the potential $\lambda,\lambda_2, \lambda_\Phi$ 
can be replaced by the physical masses $m_{\hat h}^2$, $m_{\hat s}^2$ and $\theta$:
\begin{eqnarray}
&&  \lambda     = \frac{m_{\hat h}^2 c^2_ \theta + m_{\hat \phi}^2 s^2_\theta}{2 v_{\text{EW}}^2},   \\
&&  \lambda_2   = \sin(2\theta) \frac{m_{\hat h}^2 - m_{\hat \phi}^2}{4 v_{\text{EW}} v_{\Phi}}, \label{eq:lam2} \\
&&  \lambda_\Phi = \frac{m_{\hat h}^2 s^2_ \theta + m_{\hat \phi}^2 c^2_\theta}{2 v_{\Phi}^2}. 
\label{eq:lams}
\end{eqnarray}
In summary, we have a total of 7 free parameters: 
\begin{equation}
  v_{\Phi}, \quad m_{\hat{\phi}}, \quad m_S, \quad \lambda_S, \quad \theta, \quad \lambda_1, \quad \lambda_3 ,
\end{equation}
where $\lambda_1$ and $\lambda_3$ are relevant for the DM phenomenology.

Before proceeding to study EWPT and DM phenomenology, we discuss constraints on the mixing angle $\theta$ from Higgs measurements and electroweak precision measurements. 
Couplings of the SM-like Higgs to all SM particles are rescaled by the factor $c_\theta$ due to the mixing.  
As a result,  signal rates, the ratio of Higgs measurements relative to the SM-like Higgs measurements, equal to $c_\theta^2$. 
Performing a universal Higgs fit to the data given by the ALTAS and CMS collaborations, one has $|\theta|\leq 0.526$ at the $95\%$~CL~\cite{Chao:2016vfq}. 
It was shown in Ref.~\cite{Profumo:2014opa} that the constraint of electroweak precision is closely related to the mass of $\hat \phi$,  which is weaker than that of Higgs measurements for a light $\hat \phi$ and turns to be much strong as $m_{\hat \phi}$ gets heavier.
For constraints of vacuum stability and perturbativity, we refer the reader to Refs.~\cite{Gonderinger:2009jp,Chao:2012mx} for detail.

\begin{figure*}[t]
  \includegraphics[width=0.4\textwidth]{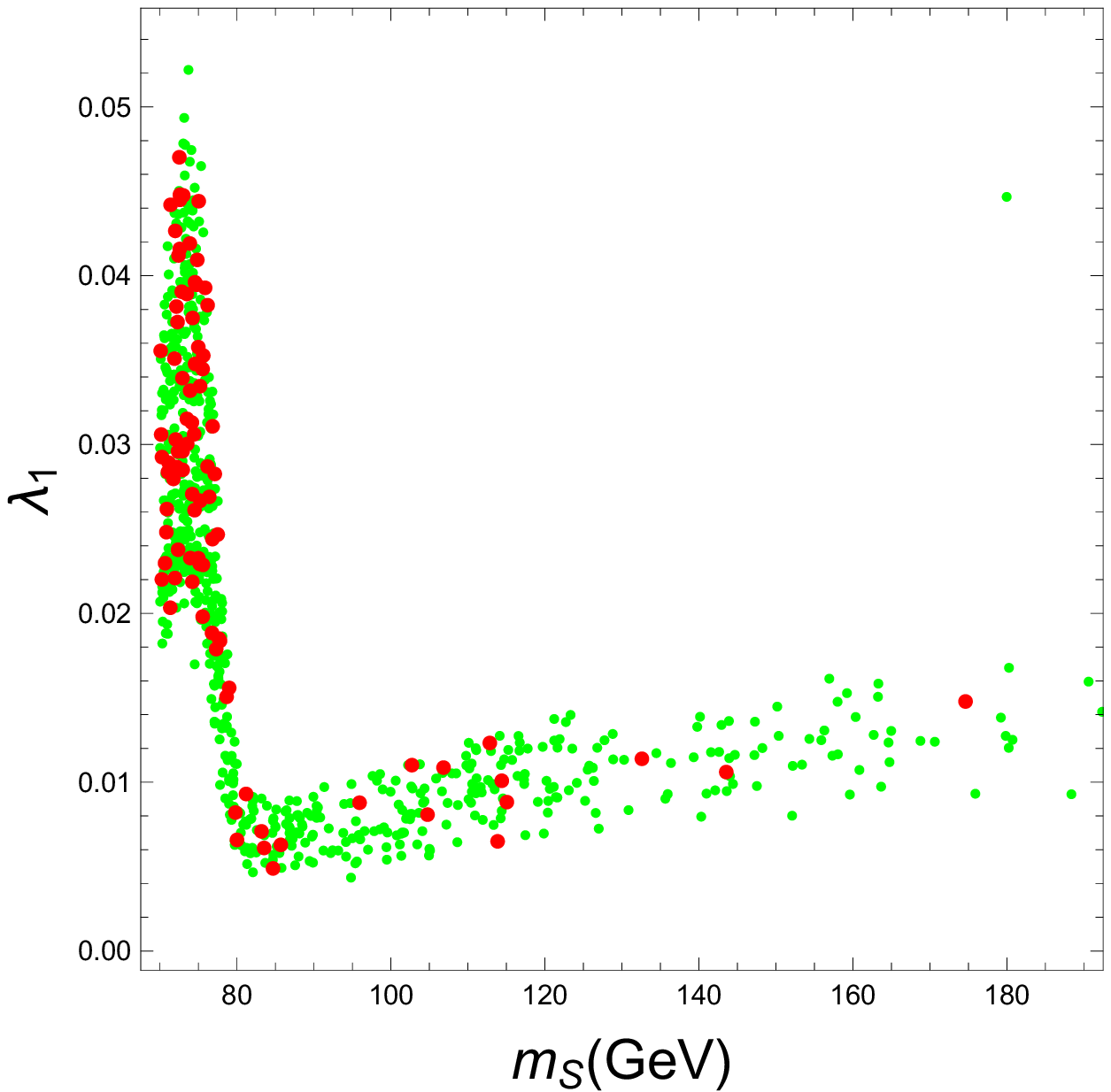} \quad \quad \quad
  \includegraphics[width=0.4\textwidth]{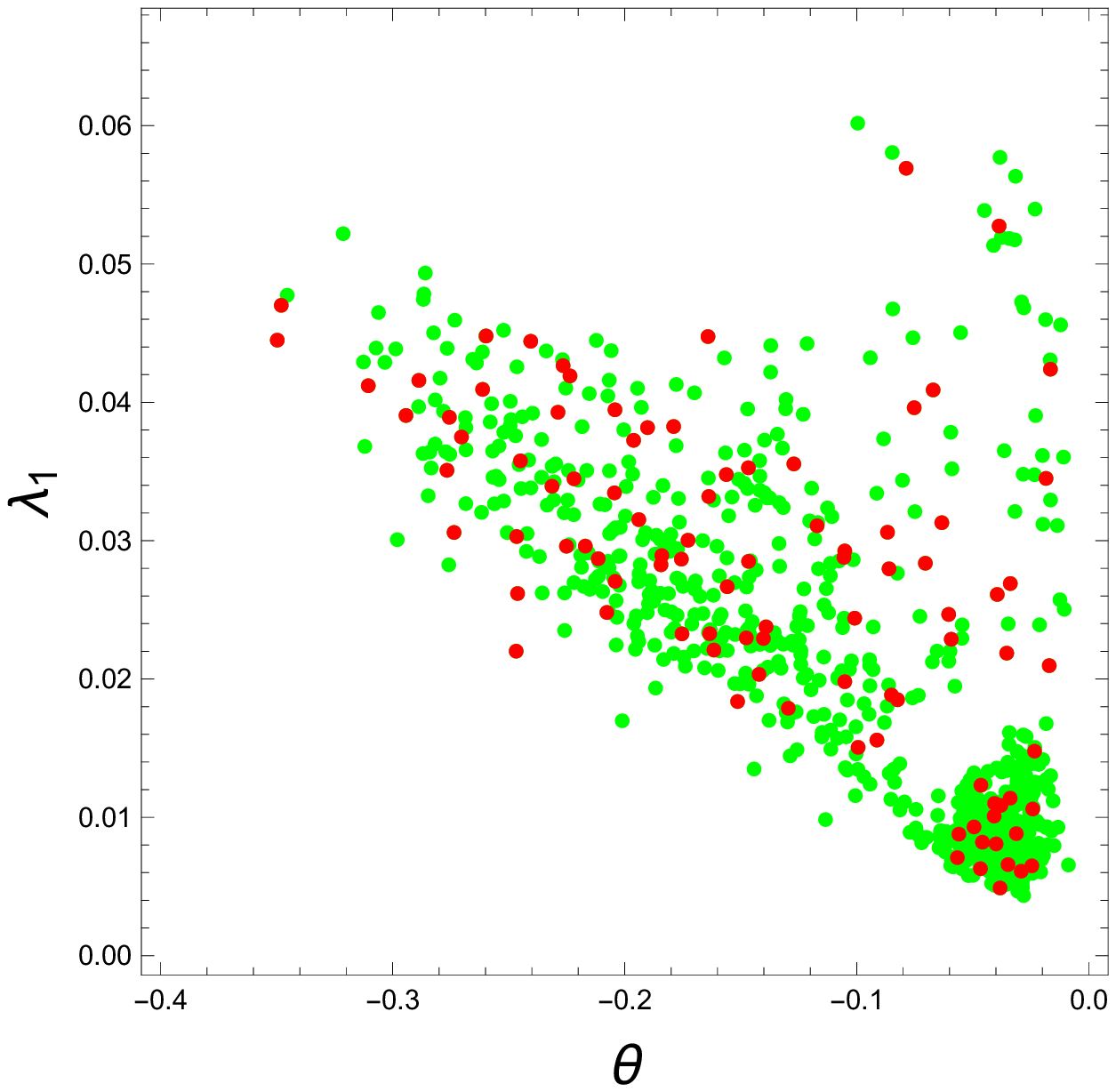}
\caption{Results from a scan over the parameter space in plane 
  $(m_S, \lambda_1)$ for the left panel and $(\theta, \lambda_1)$ for the right panel.
  Every point in these plots generates the desired EWPT pattern and leads to vanishing DM 
  nucleon scattering cross section at tree level. The green points
  further gives DM relic density within the interval $\Omega_c h^2\in (0.03, 0.12)$.
  The red points, in addition to generating relic density falling into this interval, also 
  satisfy the strong first order EWPT condition $v_h(T_C)/T_C \gtrsim 1$.
\label{fig:scan}
}
\end{figure*}

\section{\label{sec:dm}Dark Matter}
The fact that about $26.8\%$ of the universe is made of DM, whose relic abundance is about 
$\Omega_c h^2=0.1189$~\cite{Ade:2015xua}, has been well established, while the nature of the DM is still unknown. 
The Higgs potential in Eq. (\ref{potential}), has $Z_2\times Z_2$ discrete symmetry for $S$ and $\Phi$ respectively.  At the zero temperature, $\Phi$ gets non-zero VEV and thus its corresponding $Z_2$ is broken. 
On the other hand, we require that the global minimum at $T=0$ forbids a nonzero VEV in the $s$ direction.
As a result $s$ can serve as a DM candidate, which interacts with the SM particles via the Higgs 
portal interactions. 
The couplings between the DM and the physical scalars are
\begin{eqnarray}
  &&s^2\hat h^{~~}~~:    2 \lambda_3 v_\phi s_\theta +\lambda_1 v_{\text{EW}} c_\theta \; ,   \label{dd1}  \\
  &&s^2 \hat \phi^{~~}~~ : 2\lambda_3 v_\phi c_\theta -  \lambda_1 v_{\text{EW}} s_\theta  \; ,    \label{dd2} \\
&&s^2  \hat h^{2}~~ : {1\over 2 }  \lambda_1  c_\theta^2  +  \lambda_3  s_\theta^2  \; ,      \\
&&s^2 \hat \phi^2 ~~: {1\over 2 } \lambda_1 s_\theta^2 + \lambda_3 c_\theta^2  \; ,      \\
&&s^2 \hat h^{~} \hat \phi :   c_\theta s_\theta (2 \lambda_3-\lambda_1) \; . \label{interactions}
 \end{eqnarray}
The relic density of the DM is governed by the Boltzmann equation: 
\begin{eqnarray}
\dot{n} + 3 H n = -\langle  \sigma v \rangle (n^2 -n_{\rm EQ}^2) \; ,
\end{eqnarray} 
where $H$ is the Hubble constant, $\langle \sigma v \rangle$ is the thermal average of the reduced annihilation cross section. 
For our model, the DM can annihilate into $\bar{f}f$,  $ZZ/WW$, $\hat h \hat h$, 
$\hat \phi \hat \phi$  and $\hat h  \hat \phi$ final states  via s-channel Higgs mediations, where di-scalar processes also receive contribution from four point interactions in Eq.~(\ref{interactions}) .
Compared with the most simple Higgs portal, here there are more annihilation channels for a heavy DM.
Couplings in Eqs. (\ref{dd1}) and (\ref{dd2}) are both relevant to the DM direct detections,  while the contribution of $\hat \phi$ mediated processes is suppressed by the factor $s_\theta^2$ and thus is negligible for a small $\theta $. 
For numerical simulations we implement the model in  LanHEP~\cite{Semenov:1996es,Semenov:2010qt} and use  MicrOMEGAs~\cite{Belanger:2001fz,Belanger:2013oya} to calculate the  relic density.

For direct detections, the spin independent scattering cross section off nucleon is given by
\begin{eqnarray}
\sigma_n = {\mu^2  m_n^2\over  \pi  v_{\rm EW}^2  m_S^2 } \left| { c_\theta a_{\hat h} \over  m_{\hat h}^2} - {s_\theta a_{\hat \phi } \over m_{\hat \phi}^2 } \right|^2   \left( {2\over 9} + {7\over 9}\sum_{q=u,d,s} f_{T_q}^n \right)^2
\end{eqnarray}
where $a_{\hat h}$, $a_{\hat \phi}$ are effective couplings in Eqs.~(\ref{dd1})~and~(\ref{dd2}) respectively and $f^n_{T_q}$ are nucleon form factors for light quarks~\cite{Cheng:2012qr}. 
The inclusion of $\Phi$ that mixes with the SM Higgs can lead to a complete cancellation in contributions to $\sigma_n$ at tree level. 
This occurs when the effective coupling of DM with quarks vanishes, that is, when the expression in $|\cdots|$ vanishes. 
Imposing this condition would allow us to eliminate one more parameter $\lambda_3$,
\begin{eqnarray}
  \lambda_{3} = \frac{v_{\text{EW}} \lambda_1(m_{\hat{h}}^2 \tan\theta + m_{\hat{\phi}}^2 \cot\theta)}
  {2 v_{\Phi}(m_{\hat{h}}^2 - m_{\hat{\phi}}^2)} ,
\end{eqnarray}
which will be adopted in the following numerical analysis for simplicity.

For a detailed survey of the parameter space, we perform a parameter scan and show the results in 
Fig.~\ref{fig:scan} in the plane $(m_S,\lambda_1)$ (left panel) and in the plane 
$(\theta, \lambda_1)$ (right panel). 
In these plots, the green points give the relic density within the range 
$\Omega_c h^2 \in (0.03,0.12)$ allowed by the current results from Planck~\cite{Ade:2015xua}, 
in addition to being able to induce a two-step EWPT pattern as discussed later.
From the plot in the left panel, we can see the coupling 
$\lambda_1$ that can give a relic density within this range is of magnitude $(0.005,0.02)$ for a relatively
heavy DM, while it can be larger $\approx 0.06$ for a relatively light DM. 
For the right plot, the mixing angle is mostly negative since most of the scanned points are for $m_{\hat{\phi}} > m_{\hat{h}}$ and 
therefore a negative $\theta$ is needed to give a positive $\lambda_2$ according to Eq.~\ref{eq:lam2}. 
The red points in both plots also give a strongly first order EWPT condition $v_h(T_C)/T_C \gtrsim 1$.

\begin{figure}[t]
  \includegraphics[width=\columnwidth]{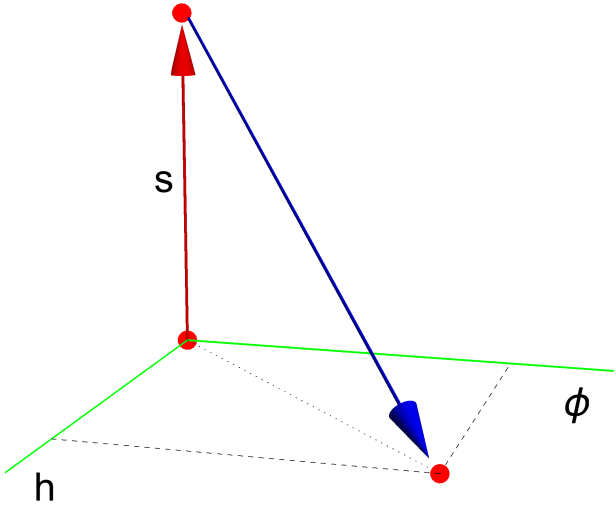} 
\caption{
This figure gives an illustrative picture of the two step EWPT
with the first step in the $s$ direction and the subsequent one from $s$ direction to the $(h,\phi)$ 
direction. 
\label{fig:EWPT-Pattern}
}
\end{figure}

\section{Electroweak Phase Transition\label{sec:ewpt}}
The barrier between the electroweak symmetric and broken phases emerges at the tree-level in our two-step phase transition scenario. 
As a result, the effective potential, which includes the standard one-loop $T\neq 0$ corrections but only retain the leading terms( thermal mass terms) in the high $T$ expansion so as to avoid problems relating to the gauge dependence, can be written as
\begin{eqnarray}
V_{\rm eff}^{} &=&  -{1\over 2 }(\mu_\Phi^2 -\Pi_\phi ) \phi^2 - {1\over 2 }(\mu_S^2  -\Pi_s) s^2 \nonumber \\ &&
 -{1\over 2 } (\mu^2 -\Pi_h) h^2 + {1\over 4} \lambda h^4+ {1\over 2 }\lambda_1^{} s^2 h^2+ {1\over 2 }\lambda_2^{} \phi^2  h^2 
 \nonumber \\ 
 && +\lambda_3^{} s^2 \phi^2   + {1\over 4 }\lambda_\Phi \phi^4  + {1\over 4}\lambda_S^{} s^4  \label{thermal}
 ,
\end{eqnarray}
in terms of background fields $h,~\phi$ and $s$.

Given Eq. (\ref{thermal}), one can trace the evolution of the universe phase as temperature drops and our
desired pattern of EWPT is illustrated in Fig.~\ref{fig:EWPT-Pattern}.
At sufficiently high temperature, the universe sits at the global minimum $(0,0,0)$ where the electroweak 
symmetry is restored. 
As temperature drops to $T_s$, a minimum develops in the $s$ direction and due to an 
absence of a barrier with the one at the origin, the universe transits to this minimum through a second 
order phase transition. 
As $T$ continues decreasing to $T_h$, a second minimum develops in the $(h,\phi)$ direction but its free energy is initially higher than the one in the $s$ direction. 
As $T$ further decreases, the free energy of the minimum in the $(h,\phi)$ direction  drops 
faster than the one in the $s$ direction, and at the critical temperature $T_C$, these two minima become degenerate. 
Slightly below $T_C$, the universe makes a second transition to the global minimum in the $(h,\phi)$ 
direction. 
Due to the existence of the barrier between these two minima, the transition may occur as a 
first order EWPT by a tunneling process and proceeds through nucleations of electroweak 
bubbles~\cite{Coleman:1977py,Linde:1980tt,Linde:1981zj} which expand, collide and coalesce, leaving eventually the universe in the electroweak broken phase.
To avoid washing out of the generated baryons by the sphaleron process inside the electroweak bubble,
the sphaleron process needs to be sufficiently suppressed inside the bubble and this translates into the following
generally adopted criterion~\cite{Shaposhnikov:1986jp,Shaposhnikov:1987tw,Cline:2006ts}:
\begin{eqnarray}
{v_h(T_C) \over T_C}  \gtrsim 1 \; . 
\label{eq:strength}
\end{eqnarray}
Note that due to the reflection symmetry of the effective potential under $h\rightarrow -h$,
$\phi \rightarrow -\phi$ or $s\rightarrow -s$, there are identical phase structures and we focus on the 
region with $h,\phi,s>0$.
%
%

\begin{figure*}[t]
  \includegraphics[width=0.44\textwidth]{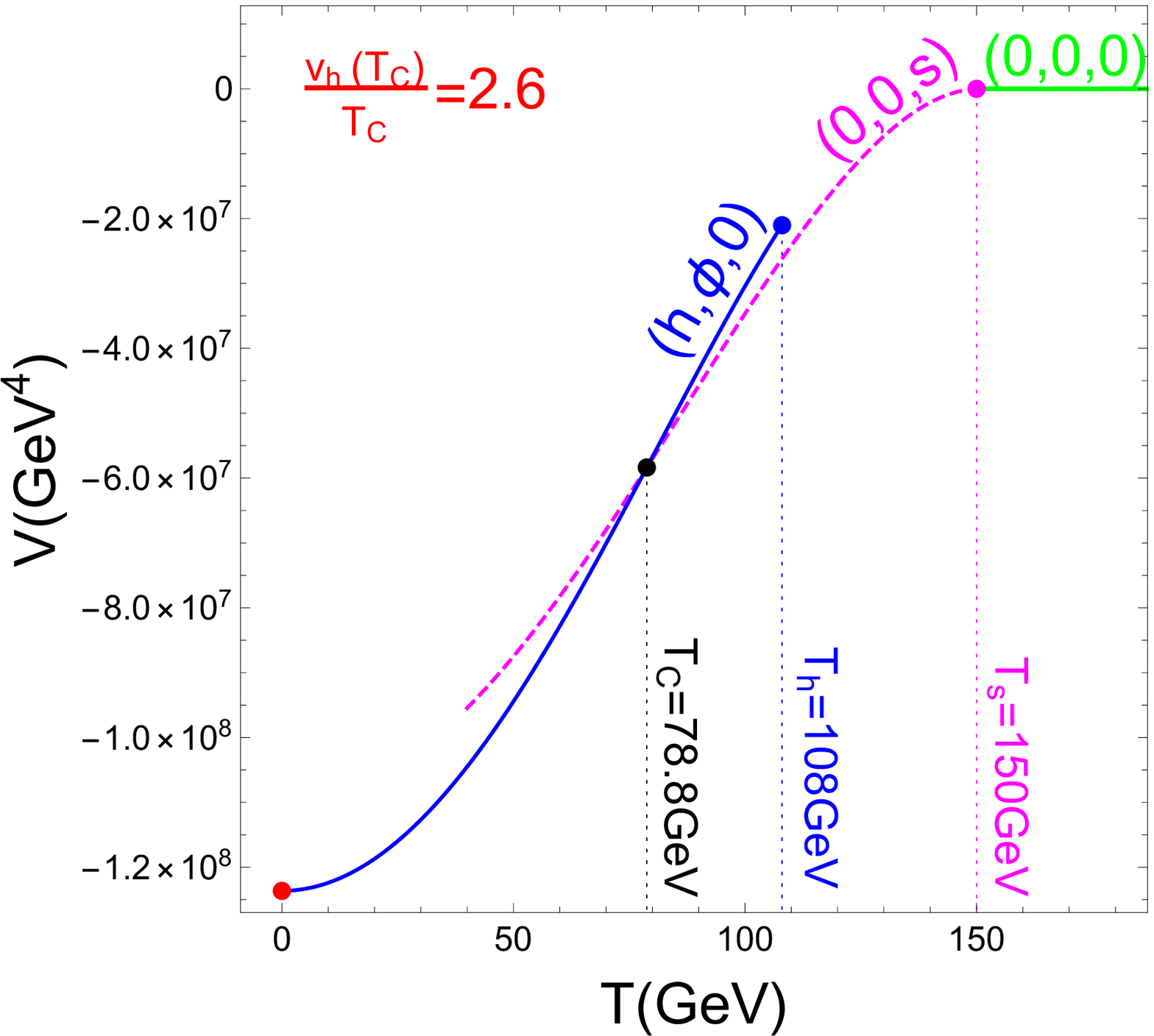} \quad \quad \quad
  \includegraphics[width=0.4\textwidth]{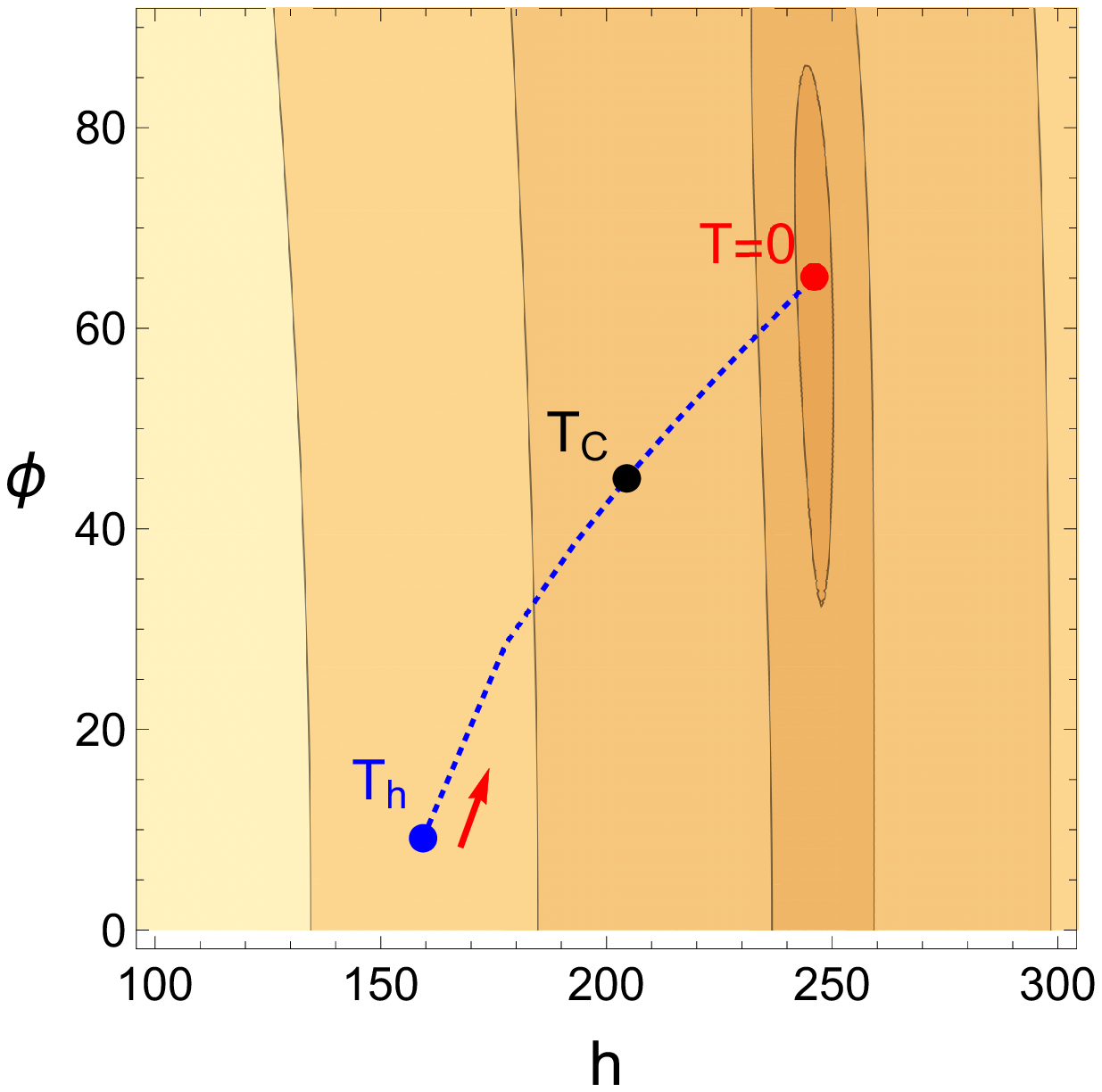}
\caption{
The left panel shows the evolution of $V$ at the two minima in $s$(blue) and in $(h,\phi)$(magenta dashed) 
directions as $T$ drops from right to left. The right panel shows the tracks of the minimum 
(blue dotted line) in the $(h,\phi)$ direction for the thermal history with the contours denoting the 
values of $V$ at $T=0$. In these plots, the magenta, blue, black and red dots represent 
the temperatures at $T_s=150\text{GeV}$, $T_h=108\text{GeV}$, $T_C=78.8\text{GeV}$ and $T=0$ respectively.
The parameters are chosen as: $v_{\Phi} = 65\text{GeV}$, $m_{\hat{\phi}}=82\text{GeV}$, 
$m_S=71\text{GeV}$, $\lambda_S=0.015$, $\theta=0.12$, $\lambda_1=0.046$ and $\lambda_{3} = 0.57$
which gives $v_h(T_C)/T_C = 2.6$.
\label{fig:EWPT}
}
\end{figure*}

The VEV of the SM Higgs at the finite temperature can be written as
\begin{eqnarray}
v_h(T) = \sqrt{ v_{\rm EW  }^2+ {\lambda_2 \Pi_\phi - \lambda_\Phi \Pi_h  \over \lambda \lambda_\Phi -\lambda_2^2} } ,
\end{eqnarray}
%
%
The critical temperature $T_C$ can be calculated analytically from the following equation: 
\begin{eqnarray}
{ \lambda (\mu_\Phi^2 -\Pi_\phi )^2   -2 \lambda_2 (\mu^2 -\Pi_h ) (\mu_\Phi^2 -\Pi_\phi) + \lambda_\Phi (\mu^2 -\Pi_h)^2 \over \lambda \lambda_\Phi -\lambda_2^2 } \nonumber \\
={ (\mu_S^2 -\Pi_s)^2  \over \lambda_S} \; . \hspace{6.2cm}   \nonumber 
\end{eqnarray}
where $\mu^2$, $\mu^2_{\Phi}$ and $\mu^2_S$ can be written in terms of physical parameters. 

As a concrete example, we show in the left panel of Fig.~\ref{fig:EWPT}, the evolution of $V$ at these 
two minima as the temperature drops from right to the left, for the parameter choice
$v_{\Phi} = 65\text{GeV}$, $m_{\hat{\phi}}=82\text{GeV}$, 
$m_S=71\text{GeV}$, $\lambda_S=0.015$, $\theta=0.12$, $\lambda_1=0.046$ and $\lambda_{3} = 0.57$. 
Here the horizontal green line on the far right 
denotes the symmetric phase at high temperature when the universe sits at the origin, the magenta dashed 
line represents the minimum in the $s$ direction 
and the blue line is the minimum in the $(h,\phi)$ direction which eventually evolves to the electroweak 
minimum at the zero temperature labeled by a red dot. 
The $s$ direction phase appears at $T_s=150\text{GeV}$ continuously away from the origin while 
the $(h,\phi)$ direction phase starts from $(159\text{GeV},9.15\text{GeV},0)$ at $T_h = 108\text{GeV}$ above which 
it is a saddle point. The $T$ at which the blue and magenta lines intersect is the critical 
temperature $T_C=78.8\text{GeV}$ and is labeled by a black dot. 
At $T_C$, the two minima are 
$(205\text{GeV},45\text{GeV},0)$ and $(0,0,352\text{GeV})$. 
For more details on the evolution 
of the minimum in the $(h,\phi)$ direction, we plot in the right panel the track of this minimum in 
blue dotted line for the whole thermal history with here the red arrow denoting the direction 
as temperature drops. The contours in this plot gives an measure of the value $V$ 
at $T=0$ and the electroweak vacuum labelled by the red point sits at the deepest location.
For this parameter choice, the strongly first order EWPT criterion~Eq.(\ref{eq:strength}) is achieved 
since $v_h(T_C)/T_C=2.6$.

A more comprehensive survey of the model is given by a scan over the parameter space as shown in 
Fig.~\ref{fig:scan} where all plotted points give above described pattern of EWPT by 
imposing various conditions during the scan. These conditions include: 
(i) there are two minima in the field space with one in the $s$ direction and the other in $(h,\phi)$ 
direction; 
(ii) the electroweak minimum needs to be lower than the one in the $s$ direction at $T=0$ should the $s$ 
direction minimum persist at $T=0$;
(iii) the minimum in the $s$ direction occurs earlier than the one in the $(h,\phi)$ direction, that is,
$T_s > T_h$. 
After imposing these conditions, the points are further filtered to give a relic density in the 
range $(0.03,0.12)$ with the remaining plotted using green color. 
At this step, these points can give a 
two-step EWPT and a sizable amount of relic density but the second step EWPT is not necessarily strongly 
first order. 
As such, we further calculate the critical temperature $T_C$
and $v_h(T_C)$ corresponding to the black intersection point of the blue and magenta lines for each above 
green point. The points which satisfy the condition in Eq.~(\ref{eq:strength}) are shown
with red color in Fig.~\ref{fig:scan} . 
So we can see there are sufficient parameter space in this model where the desired strongly
first order EWPT pattern can be realized. 
We note in passing that the second order phase transition for the first 
step could also be strongly first order, should a more complete effective potential be adopted.

\section{Gravitational Waves\label{sec:gw}}
The calculation of the GWs generated during the second step EWPT needs a numerical 
analysis of the tunneling process at finite temperature and in particular involves solving the critical bubble profiles.
As mentioned in previous section, 
after the universe cools down to a temperature below $T_C$, the second transition from 
$(0,~0,~v_s)$ to the true 
vacuum $(v_h, ~v_\phi,~0)$ takes place by the nucleation of true vacuum bubbles. 
This tunneling rate per unit time per unit volume reads~\cite{Moore:1995si}
\begin{eqnarray}
\Gamma \sim A(T) e^{-S_3/T} ,
\end{eqnarray} 
where $S_3$ is the Euclidean action of the critical bubble which minimizes the action
\begin{eqnarray}
S_3(\vec{\phi},T) = 4\pi \int r^2 dr \left[ \frac{1}{2} \left(\frac{d \vec{\phi}(r) }{d r}\right)^2 + V(\vec{\phi},T) \right], 
\end{eqnarray}
and the prefactor  $A(T)$ is roughly of  ${\cal O} (T^4)$, whose 
precise evaluation needs integrating out fluctuations around the bounce solution~\cite{Dunne:2005rt}.
From this rate formula, the bubble nucleation temperature $T_n$ is defined as the probability for a 
single bubble to be nucleated within one horizon volume being ${\cal O}(1)$, i.e.
\begin{eqnarray}
\int_0^{t_n}  {\Gamma V_{H} (t)}  dt= \int_{T_n}^{\infty} {d T \over T }  \left( 2 \zeta M_{\rm pl}  \over T  \right)^4 e^{{-S_3/T}} ={\cal O} (1) , \ 
\label{nucleation}
\end{eqnarray}
where $V_H (t)$ is the horizon volume, $M_{\rm pl}$ is the Planck mass and $\zeta \sim 3 \times 10^{-2}$. 
This equation implies numerically $S_3(T_n)/T_n\approx 140$~\cite{Apreda:2001us} and serves as 
our definition of $T_n$.
%
\begin{figure*}[t]
  \includegraphics[width=0.4\textwidth]{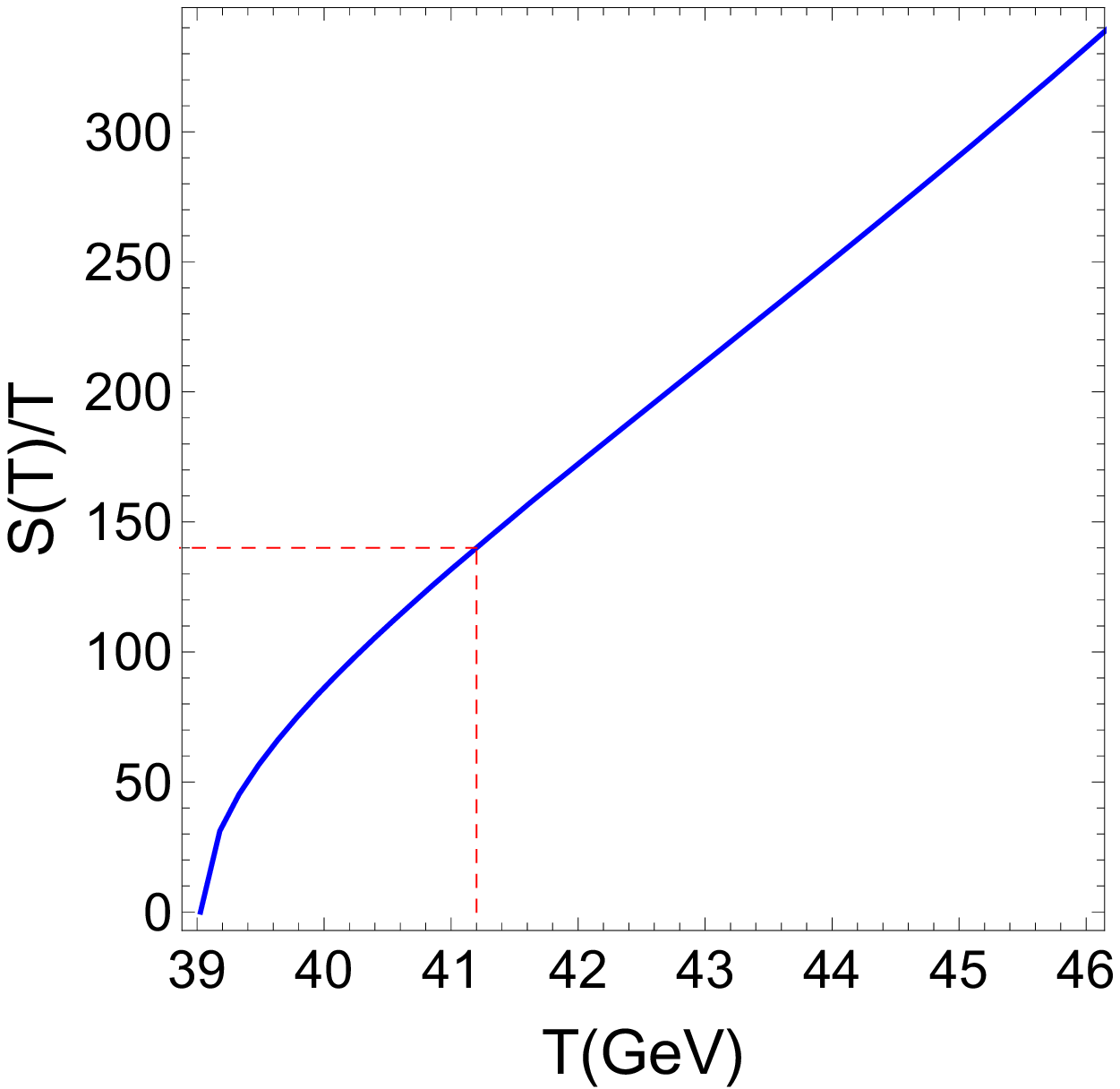} \quad \quad \quad
  \includegraphics[width=0.42\textwidth]{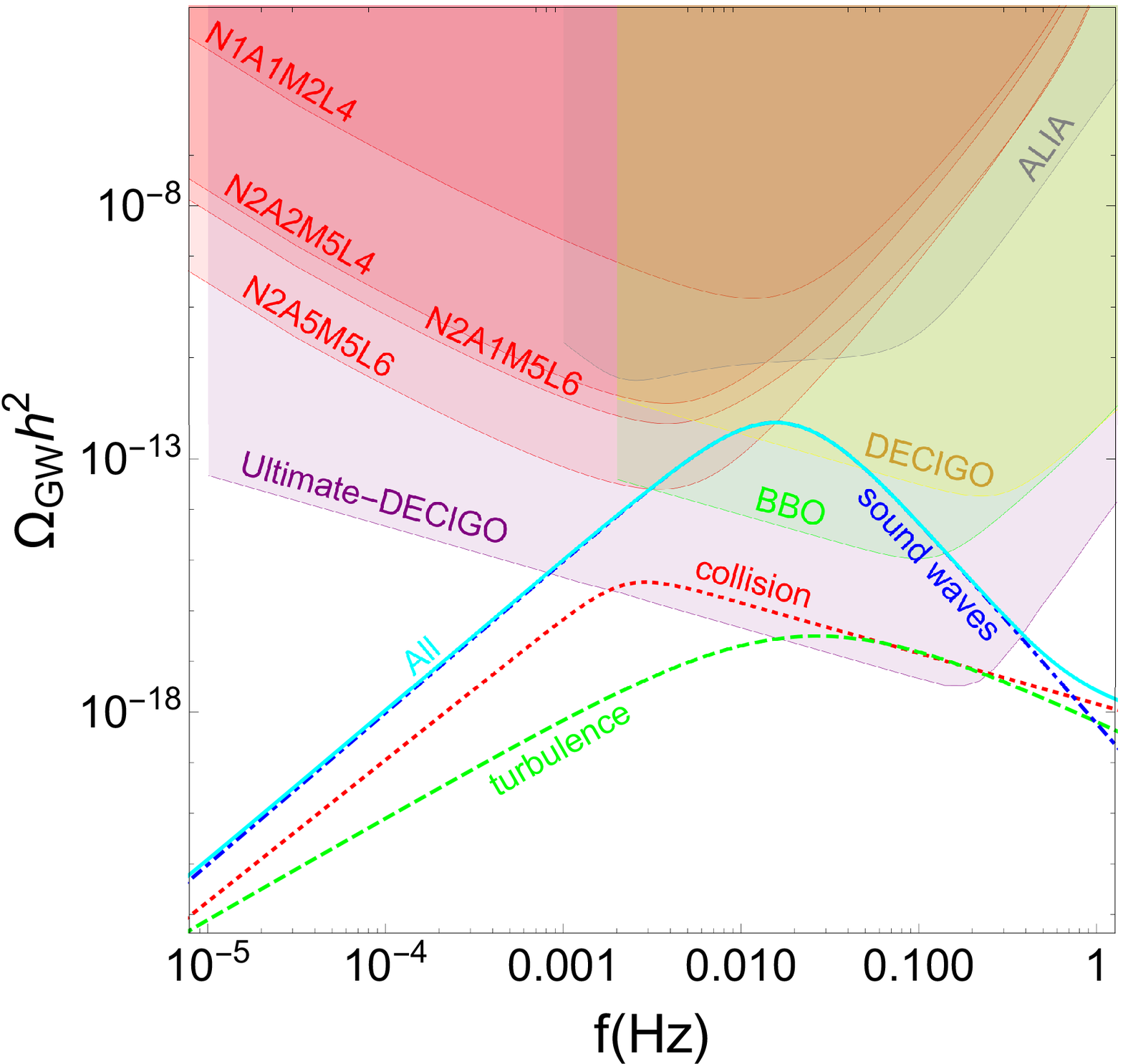}
  \caption{
    The left panel shows changes of $S(T)/T$ in the neighborhood of $T_n$ and the right panel shows 
    the
    GWs generated during the first order EWPT as a function of frequency
    from three sources: sound waves(blue dotdashed line), collision(red dotted line), 
    turbulence(green dashed line) and total contribution(cyan solid line). 
  The color shaded regions fall within the experimental sensitivities of
    eLISA(four configurations with notation NiAjMkLl), ALIA(gray), BBO(green), DECIGO(yellow) and 
Ultimate-DECIGO(purple).
\label{fig:gw}
  }
\end{figure*}
From the bounce solutions, two other important parameters $\alpha$ and $\beta$, that are directly 
relevant for the calculation of GWs, are defined by:
\begin{eqnarray}
\alpha \equiv {\rho_{\rm vac} \over \rho^*_{\rm rad}} \; , \hspace{1cm} \beta \equiv H_nT_n \left.{d S_3 \over d T} \right |_{T_n} \; , 
\end{eqnarray}
where $\rho_{\rm vac} $ is the vacuum energy density released in the phase transition, $H_n$ is the Hubble parameter at $T_n$, $\rho_{\rm rad}^* =g_* \pi^2 T_n^4 /30$ with $g_*$ the relativistic 
degrees of freedom in the plasma at 
$T_n$.   
A small $\beta/H_n$ will trigger strong phase transition and consequently a significant stochastic background of GWs.

The observable characterizing the GW background is the energy spectrum $h^2 \Omega_{\text{GW}}(f)$ 
and is given by~\cite{Apreda:2001us}
\begin{eqnarray}
h^2 \Omega_{\text{GW}}(f)=\frac{h^2}{\rho_c}\,\frac{d\rho_{\rm gw}}{d\log f}\, ,
\end{eqnarray}
where $\rho_{\rm gw}$ is the energy density of GWs with $f$ its frequency and 
$\rho_c$ is the critical energy density today.
The sources of stochastic GW signals arising from first order EWPT can be classified into three categories: (1) collisions of bubble walls and shocks in the plasma; (2) sound waves in the plasma after the bubble collision; (3) Magnetohydrodynamic (MHD) turbulence in the plasma~\cite{Caprini:2015zlo}. 
The total energy spectrum is given approximately by the sum of these three sources,
\begin{eqnarray}
h^2 \Omega_{\rm GW} \approx h^2 \Omega_{\rm coll}^{} + h^2 \Omega_{\rm sw}^{} + h^2 \Omega_{\rm turb}^{}  \; .
\end{eqnarray}
The GW contribution from bubble collisions can be calculated using the envelope approximation~\cite{Hawking:1982ga,Caprini:2007xq,Jinno:2016vai}, which numerically results in the following GW contribution to the spectrum~\cite{Huber:2008hg}:
\begin{eqnarray}
h^2 \Omega_{\rm coll} (f) &=& 1.67\times 10^{-5} \left({ H_n \over \beta } \right)^2 
\left( {   \kappa \alpha \over 1+ \alpha }\right)^2 \left( { 100 \over g_*} \right)^{1\over 3 }  \nonumber \\ &&  \times \left( {0.11 v_w^3 \over 0.42 + v_w^2  }\right) \left[  { 3.8 (f /f_{\rm coll})^{2.8} \over 1+ 2.8 (f/f_{\rm coll})^{3.8}} \right] ,
\end{eqnarray}
where $v_w$ is bubble wall velocity, $\kappa$ characterizes  the fraction of latent heat deposited in a thin shell and $f_{\rm coll}$ is the peak frequency produced by the bubble collisions.
At the time of the phase transition $f^n_{\rm coll} =0.62\beta /(1.8-0.1v_w+v_w^2)$, which is 
red-shifted to give the peak frequency today: 
$f_{\rm coll} = f^n_{\rm coll} \times a (T_n)/a_0=16.5\times 
10^{-6}~\text{Hz} ~(f^n_{\rm coll}/H_n)(T_n/100~{\rm GeV}) (g_*/100)^{1/6}$.
Both $v_w$ and $\kappa$ are functions of $\alpha$, which read as~\cite{Kamionkowski:1993fg}:
\begin{eqnarray}
v_w = \frac{1/\sqrt{3} + \sqrt{\alpha^2+2\alpha/3} }{1+\alpha}, \quad \quad  
\kappa \simeq \frac{0.7\alpha + 0.2 \sqrt{\alpha} }{1+0.7\alpha} .
\end{eqnarray}
The sound wave contribution of the GW intensity is numerically fitted by 
\begin{eqnarray}
h^2 \Omega_{\rm sw} (f) &=&2.65\times 10^{-6} \left(  {H_n \over \beta }\right) \left( { \kappa_v \alpha \over 1+ \alpha} \right)^2  \left( {100 \over g_* } \right)^{1\over 3} \nonumber \\&& \times v_w  \left( {f \over f_{\rm sw}} \right)^{3} \left[{7 \over 4+ 3 (f/f_{\rm sw})^2} \right]^{7/2} \; ,
\end{eqnarray}
where $\kappa_v$ denotes the faction of latent heat transformed into bulk motion of the fluid, $f_{\rm sw}$ is the peak frequency that can be given from the rescaling of its value at the phase transition, i.e. $f_{\rm sw}^{} = f_{\rm sw}^n \times a(T_n)/a_0 = (2/\sqrt{3}) (\beta/v_w) \times a(T_n)/a_0 $. 
We refer the reader to Refs.~\cite{Espinosa:2010hh,Caprini:2015zlo} for the value of $\kappa_v$ in the small and large $v_w$ limit. 
The MHD turbulence contribution to the GW spectrum can be written as
\begin{eqnarray}
h^2 \Omega_{\rm turb} (f)& = &3.35 \times 10^{-4} \left(  {H_n \over \beta }\right) \left(  { \kappa_{\rm tu } \alpha \over 1+ \alpha }\right)^{3/2}\left( {100 \over g_* } \right)^{1\over 3 } \nonumber \\ && \times v_w { ( f/f_{\rm tu})^3 \over (1+ f/f_{\rm tu})^{11/3} (1+8\pi f/h_n)} \ ,
\end{eqnarray}
where $\kappa_{\rm tu} \approx 0.1 \kappa_v$~\cite{Hindmarsh:2015qta}, $h_n$ is the Hubble parameter today, and $f_{\rm tu} \approx (3.5/2) (\beta /v_w ) \times a(T_n)/a_0$.

We use the package CosmoTransitions~\cite{Wainwright:2011kj} to solve for profiles of the 
critical bubble and the nucleation temperature for the benchmark point used earlier. 
This parameter choice gives $\Omega_c h^2 = 0.04$ constituting about 
$34\%$ of the total DM relic density. 
For this benchmark parameter point, we show in the left panel of Fig.~\ref{fig:gw} the quantity
$S(T)/T$ as a function of $T$ and we found $T_n = 41.2\text{GeV}$.
The resulting GW signals are shown in the right panel of Fig.~\ref{fig:gw} where the three 
contributions are shown with details in the caption. 
As can be seen from this figure, the blue sound wave 
contribution dominates and is almost indistinguishable from the sum of the three sources which is denoted
by the cyan line. 
For a comparison with experiment, we show firstly four sensitive regions corresponding to four configurations of the eLISA detector. These four regions are labelled in the format 
NiAjMkLl and are plotted as the red shaded regions at the top following conventions of Ref.~\cite{Caprini:2015zlo,Klein:2015hvg}. We can see for this parameter set, the GWs can be detected by the configuration N2A5M5L6 while unreachable by the others. 
We further add the sensitivitive regions of several other proposed decihertz GW experiments: 
Advanced Laser Interferometer Antenna (ALIA)~\cite{Gong:2014mca}\footnote{The ALIA program is the upgraded version of the Chinese Taiji Program in space of gravitational wave physics.}, BBO, DECIGO and Ultimate-DECIGO~\cite{Kudoh:2005as}. The data are taken from Ref.~\cite{Moore:2014lga,Yagi:2011wg,Artymowski:2016tme} and are plotted as gray, green, yellow and purple regions respectively~\footnote{The ALIA, BBO and DECIGO data are taken from the website~\url{http://rhcole.com/apps/GWplotter/} where these
sensitivity curves are collected in one place. We also checked that these curves 
are consistent with those used in 
Ref.~\cite{Artymowski:2016tme}.}.
From the plot, we can see the GW signals from the benchmark point fall within the detectable ranges of BBO, 
DECIGO and Ultimate-DECIGO.
It should be noted that there might be other points in the model parameter space where the generated GWs can have much larger energy density spectrum and are 
therefore within reach of the other three eLISA configurations as well as ALIA, and give a sufficiently large DM relic density in the meanwhile.
This however needs a dedicated survey of the parameter space including all  the considerations and we leave it to future works.
For GWs from one-step EWPT in SM extended with scalar singlet(s)~\footnote{GWs can be generated from EWPT in SM extended with dimension-six operators~\cite{Huang:2016odd,Baldes:2017rcu,Addazi:2016fbj}, where no scalar singlet is needed. }, we refer the reader to Refs.~\cite{Jinno:2016knw,Huang:2016cjm,Hashino:2016xoj,Kakizaki:2015wua,Balazs:2016tbi,Jinno:2015doa} as well as references cited in these papers for detail.   
Alternatively, first order phase transition may lead to originations of primordial black holes.  They can be captured by neutron stars or astrophysical black holes~\cite{Davoudiasl:2016mwf}, resulting in GWs that may be detected by Advanced LIGO or Advanced Virgo. 

\section{Summary\label{sec:summary}}
Working in a simple model with the SM extended by a scalar singlet DM and another scalar singlet which mixes with the SM-like Higgs boson, we studied a possible connection between the DM phenomenology and the EWPT, in particular the detectability of the GW signals generated during the DM assisted  EWPT.
Through both analytical and numerical studies, we find this model may admit strongly first order two-step EWPT in certain parameter space, which may also give rise to a viable the DM relic density and a negligible direct detection cross section. 
We further exemplified, using one representative benchmark point, the discovery possibility of the EWPT with the generated GW signals during the second step EWPT and found that the GW signals can be detected by the eLISA in the configuration N2A5M5L6, BBO, DECIGO and Ultimate-DECIGO. 
This scenario can readily be generalized to other  models where DM can have nontrivial effect in the baryon number generation and is helpful in understanding and testing the origin of these two cosmological puzzles.

\section{Acknowledgements}
This work is supported by the National Natural Science Foundation of China under grant 
No.11647601, No.11690022 and No.11675243 
and also supported by the Strategic Priority Research Program of the Chinese Academy of Sciences under 
grant No.XDB23030100. Part of the results described in this paper are 
obtained on the HPC Cluster of SKLTP/ITP-CAS.

\bibliographystyle{utphys}
\begingroup\raggedright
\end{document}